\documentclass[fleqn,usenatbib]{mnras}
\usepackage{newtxtext,newtxmath}
\usepackage[T1]{fontenc}

\usepackage{ae,aecompl}
\usepackage{comment}

\usepackage{microtype}
\usepackage{amsmath}
\usepackage{graphicx}
\usepackage{lscape}
\usepackage{color}

\usepackage[normalem]{ulem}

\graphicspath{{pics/}}
\newcommand {\bc}{\begin {center}}
\newcommand {\ec}{\end {center}}
\newcommand {\be}{\begin {equation}}
\newcommand {\ee}{\end {equation}}
\newcommand {\beq}{\begin {eqnarray}}
\newcommand {\eeq}{\end {eqnarray}}
\newcommand {\ergs}{{\rm erg\ \rm s$^{-1}$}}
\newcommand {\ergscm}{{\rm erg\ \rm s$^{-1}$\ \rm cm$^{-2}$}}

\newcommand{\blue}[1]{\textcolor{blue}{#1}}

\definecolor{mypink1}{rgb}{0.858, 0.188, 0.478}

\def\flux{erg~s$^{-1}$~cm$^{-2}$}
\def\lum{erg~s$^{-1}$}

\def\eras{eRASSU\,J050810.4$-$660653}

\def\srg{{\it SRG}}
\title[New X-ray pulsar in LMC: \eras]
{First characterization of a new High Mass X-ray Binary in LMC \eras\ with \textit{SRG}/ART-XC, \textit{NuSTAR} and \textit{Swift}}

\author[A.~Salganik et al.]
{Alexander~Salganik,$^{1,2}$\thanks{E-mail: alsalganik@gmail.com} 
Sergey~S.~Tsygankov,$^{3,2}$
Alexander~A.~Lutovinov,$^{2}$
Alexander~A.~Mushtukov,$^{4,5}$
\newauthor
Ilya~A.~Mereminskiy,$^{2}$
Sergey~V.~Molkov,$^{2}$
Andrei~N.~Semena$^{2}$
\\
$^1$Department of Astronomy, Saint Petersburg State University, Saint-Petersburg 198504, Russia\\
$^2$Space Research Institute of the Russian Academy of Sciences, Profsoyuznaya Str. 84/32, Moscow 117997, Russia\\
$^3$Department of Physics and Astronomy,  FI-20014 University of Turku, Finland\\
$^4$ Astrophysics, Department of Physics, University of Oxford, Denys Wilkinson Building, Keble Road, Oxford OX1 3RH, UK\\
$^5$ Leiden Observatory, Leiden University, NL-2300RA, Leiden, The Netherlands
}
\date{Accepted 2022 June 7. Received 2022 May 23; in original form 2022 February 26}

\pubyear{2022}

\begin{document}
\label{firstpage}
\pagerange{\pageref{firstpage}--\pageref{lastpage}}
\maketitle

%%%%%%%%%%%%%%%%%%%%%%%%%%%%%%%%%%%%%%%%%%%%%%%%%%%%%%%%%%%%%%%%%%%%%%%%%%%%%%
%% Abstract, Keywords and contact details                                   %%
%%%%%%%%%%%%%%%%%%%%%%%%%%%%%%%%%%%%%%%%%%%%%%%%%%%%%%%%%%%%%%%%%%%%%%%%%%%%%%
\begin{abstract}
We report results of the first detailed spectral and temporal studies of the recently discovered Be/X-ray binary \eras\ in LMC based on the data from the {\it SRG}/ART-XC, {\it NuSTAR} and {\it Swift}/XRT instruments obtained in December 2021 - May 2022 in a wide energy range of 0.5-79 keV. Pulsations with the period of $40.5781 \pm 0.0004$~s were found in the source light curve 
with the pulsed fraction monotonically increasing with the energy. An estimate of the orbital period of $\sim38$ days was obtained based on the long-term monitoring of the system. The source spectrum can be well approximated with a power-law model modified by an exponential cutoff at high energies. The pulse phase-resolved spectroscopy shows a strong variation of spectral parameters depending on the phase of a neutron star rotation. 
We have not found any features connected with the cyclotron absorption line both in the phase-averaged and phase-resolved spectra of \eras. However, the neutron star magnetic field was estimated around {\it several} $10^{13}$~G using different indirect methods. Discovered variations of the hardness ratio over the pulse phase is discussed in terms of physical and geometrical properties of the emitting region.
\end{abstract}


%%%%%%%%%%%%%%%%%%%%%%%%%%%%%%%%%%%%%%%%%%%%%%%%%%%%%%%%%%%%%%%%%%%%%%%%%%%%%%
%% Introduction                                                             %%
%%%%%%%%%%%%%%%%%%%%%%%%%%%%%%%%%%%%%%%%%%%%%%%%%%%%%%%%%%%%%%%%%%%%%%%%%%%%%%
\section{Introduction}
\label{sec:intro}
On December 8, 2019 the eROSITA telescope on board the \textit{Spektrum-Roentgen-Gamma (SRG)} mission 
discovered a new X-ray source \eras\ in the Large Magellanic Cloud (LMC) \citep{Haberl2020}. The source was localized at RA $=05^{\rm h}08^{\rm m}10\fs4$, Dec. $=-66\degr06\arcmin53\arcsec$ with an error radius of 5.1 arcsec. The source spectrum was described by an absorbed power law with the photon index of $\Gamma \sim 1.2$ and column density of $N_{\rm H}\sim 2.1\times10^{21}~\text{cm}^{-2}$ with the flux of $3.4\times10^{-12}$ \ergscm\ in the 0.2-10.0~keV energy band. The source position was consistent with that of an early-type star, making \eras\ a High Mass X-ray Binary (HMXB). An optical spectroscopy with Southern African Large Telescope (SALT) revealed single-peaked H$\alpha$ emission line, which was found to dominate the spectrum, indicating a Be/X-ray binary nature of the system \citep{Haberl2020}. On March 29 and April 1, 2020 the source was observed by the XRT telescope onboard of the Neil Gehrels \textit{Swift} observatory. The measured flux of $(1.6 \pm 0.3) \times 10^{-12}$ \ergscm\ (0.2-10.0~keV) turned out to be only two times less than measured by eROSITA four month before \citep{Haberl2020}.

\textit{SRG}/eROSITA observed  the source again in December 2020 during the second all-sky survey and \textit{XMM-Newton} TOO observations was triggered and performed on December 17, 2020 \citep{Haberl2021}. 
The power spectrum based on the EPIC-PN detector data revealed strong X-ray pulsations with a period of 40.60254(8)~s. A detailed study of the source behaviour in the low-luminosity state based on the \textit{XMM-Newton} and \textit{SRG}/eROSITA data is presented in a separate paper by Haberl et al. (in preparation).

One year later, on December 14-15, 2021 an enhanced X-ray activity from \eras\ was detected with the Mikhail Pavlinsky ART-XC telescope on board the \textit{SRG} observatory in the harder X-ray band. The preliminary estimated source flux of $\sim$2 mCrab  ($3\times10^{-11}$ \ergscm) in the 4-12~keV energy range corresponds to the luminosity of about $10^{37}$\ergs\ at the LMC distance \citep{Haberl2021}. Follow-up \textit{NuSTAR} TOO observations confirmed the presence of the pulsations in the \eras\ light curve, making it a new X-ray pulsar (XRP) in HMXB.  

In this paper, we provide a detailed analysis of the spectral and temporal properties of the \eras\ based on the {\it SRG}/ART-XC, \textit{Swift}/XRT and \textit{NuSTAR} data obtained during the December 2021 - January 2022 monitoring campaign.

\section{Data analysis}
\label{sec:data}
\subsection{{\it NuSTAR} observatory}
\textit{NuSTAR} consists of two identical co-aligned X-ray telescope modules, each equipped with focal plane detectors called FPMA and FPMB \citep{Harrison2013}. The telescope's optics provide X-ray data in a wide 3–79~keV energy range with an angular resolution of 18 arcseconds (FWHM) and a uniquely high sensitivity at high photon energies.

The effective exposure time of the utilized \textit{NuSTAR} observation is 56 ks (ObsID 90701342002). We extracted spectra and light curves using {\sc nuproducts} procedure provided by the {\sc NuSTARDAS} pipeline in accordance with the data analysis manual.\footnote{\url{https://heasarc.gsfc.nasa.gov/docs/nustar/analysis/nustar_swguide.pdf}} Data for analysis were extracted from a 40 arcsec radius circle. To subtract the background, we took data from the region with a radius of 120 arcsec. Background-subtracted spectra were binned with the "optimal" algorithm implemented in the {\sc ftgrouppha} utility of the {\sc ftools} package \citep{Kaastra2016}. Background-subtracted light curves from two modules were added using the {\sc lcmath} utility and were barycenter-corrected using the {\sc barycorr} utility. For the analysis, we used {\sc heasoft} package version 6.29 and {\sc CALDB} version 20211202.

\subsection{{\it Swift} observatory}

\begin{figure}
    \centering
    \includegraphics[width=0.99\columnwidth]{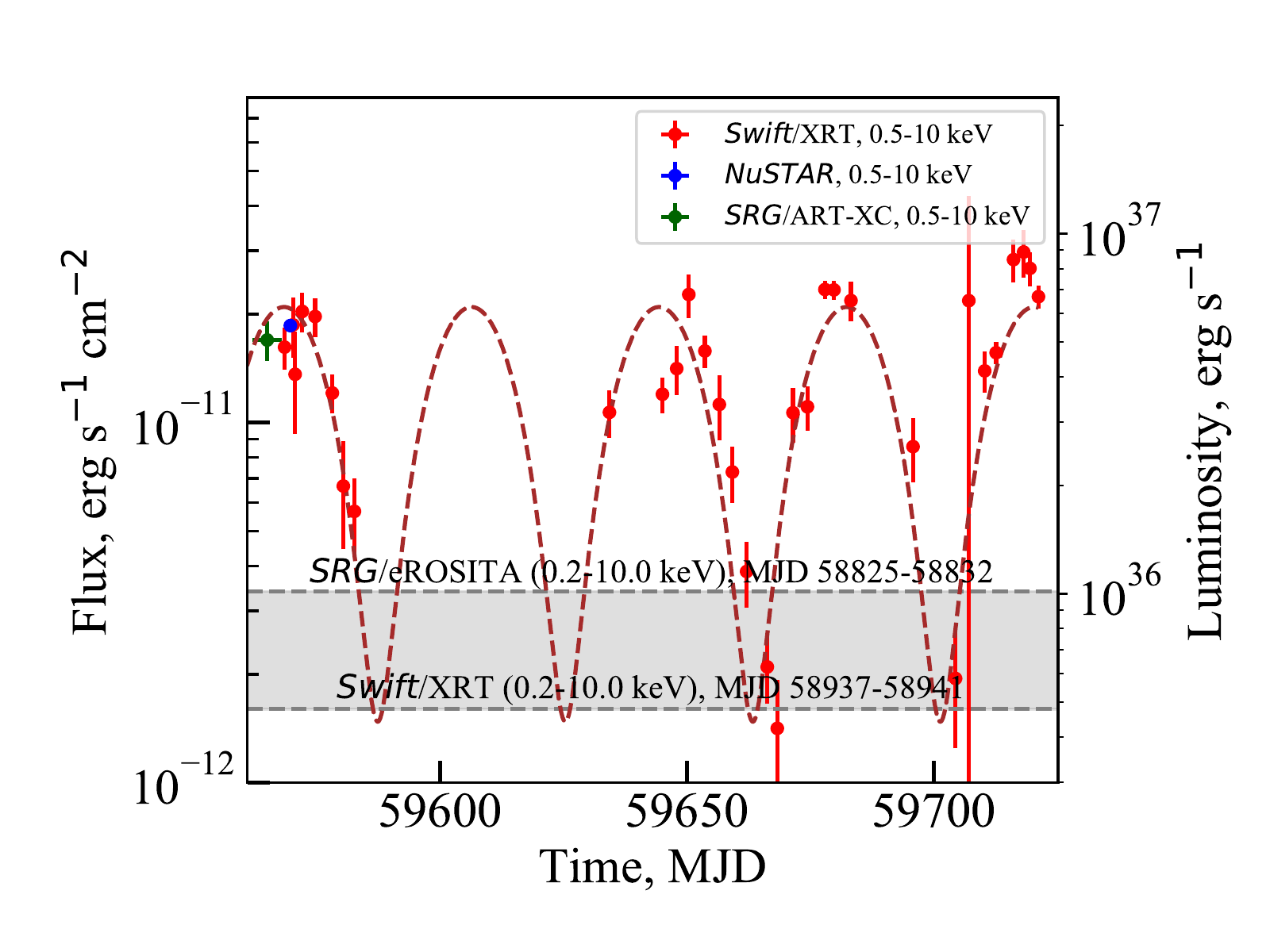}
 	\caption{Light curve of \eras\ based on the {\it Swift}/XRT telescope monitoring campaign in the 0.5-10~keV energy range (red points), {\it NuSTAR} (blue), {\it SRG}/ART-XC (green). Results of eROSITA and {\it Swift}/XRT from \citet{Haberl2020} are shown with the gray dashed lines as well as the range between them. The dotted line shows the approximation of the light curve by a sinusoid with period of $\sim38$ days.}
	\label{fig:lightcurve}
\end{figure}

To study the evolution of the \eras\ flux (see Fig.~\ref{fig:lightcurve}) and spectrum during the outburst, we triggered a monitoring campaign consisting of 32 observations (ObsID 00013299003-7,9-11, 35-58, see Table~\ref{table:observations})
with the XRT telescope \citep{Burrows2005} on board the Neil Gehrels \textit{Swift} Observatory \citep{Gehrels2004}, covering MJD 59568-59722. Unfortunately, due to the switch of the \textit{Swift} observatory into the safe mode, there were no observations between 59583 and 59634 MJD. All observations were made in the Photon Counting (PC) mode. The spectra were extracted using the data analysis software\footnote{\url{https://www.swift.ac.uk/user_objects/}} \citep{Evans2009} provided by the UK Swift Science Data Center. 
They were rebinned to have at least one count per energy bin and W-statistics was applied \citep{Wachter1979}. 
The \textit{Swift}/XRT spectra were fitted in the {\sc XSPEC} 12.12.0 package \citep{Arnaud1996} with a simple absorbed power-law model (the hydrogen column density $N_{\rm H}$ was fixed at $0.27\times10^{22}$~cm$^{-2}$ for all \textit{Swift}/XRT observations, the best-fit value obtained from the broadband spectral analysis, see Sec.~\ref{sec:spectrum}). All errors are given at the 1$\sigma$ confidence level if not specified otherwise. All luminosities within the paper were calculated assuming the  distance to the LMC of $49.97\pm1.3$~kpc \citep{2013Natur.495...76P}.

%======================================================
%======================================================

\subsection{\textit{SRG}/ART-XC}

The \textit{Spectrum Roentgen Gamma (SRG)} observatory \citep{Sunyaev2021} consists of two X-ray telescopes: the Mikhail Pavlinsky ART-XC operating in the 4–30 keV energy range \citep{Pavlinsky2021} and eROSITA operating in the 0.2–10 keV band \citep{Predehl2021}. 

The ART-XC observed the position of the \eras\, during the ongoing all-sky survey from 14 to 19 December, 2021, for a net (i.e. not corrected for vignetting) exposure of $\sim0.55$ ks. The source was clearly detected at an average flux of $(1.6\pm0.2)\times10^{-11}$ \flux\, in the 4$-$12 keV energy band.

Assuming the same spectral shape during the \textit{NuSTAR} observation (see Sec.~\ref{sec:spectrum}), scaling factor of 1.06 was used to convert ART-XC fluxes into the 0.5-10 keV band, $F_{\rm ART, 0.5-10\,keV } = 1.06 \times F_{\rm ART, 4-12\,keV }$. Data for spectral and timing analysis were reduced using the {\sc artproducts v0.9} pipeline  with the {\sc CALDB} version 20200401. The source spectrum and light curve were extracted with the {\sc artproducts v0.9} pipeline in the low photon statistic mode from the 2\arcmin\ vicinity of the source position.

%=====================================================
\begin{table}
\caption{Observations of \eras.}
\label{table:observations} 
\begin{tabular}{|l|ccc|} 
\hline 
ObsID & Exposure, ks & Count rate\blue{$^{\rm a}$} & Bkg count rate\blue{$^{\rm b}$} \\ 
\hline 
\multicolumn{4}{|c|}{\textit{Swift}/XRT} \\ 
00013299003 & 0.99 & 1.19 & 0.02 \\ 
00013299004 & 0.31 & 1.83 & 0.06 \\ 
00013299005 & 0.14 & 1.33 & 0.07 \\ 
00013299006 & 0.80 & 1.70 & 0.03 \\ 
00013299007 & 0.90 & 1.55 & 0.01 \\ 
00013299009 & 1.20 & 1.13 & 0.02 \\ 
00013299010 & 0.27 & 0.69 & 0.03 \\ 
00013299011 & 0.69 & 0.62 & 0.02 \\ 
00013299035 & 0.97 & 0.94 & 0.05 \\ 
00013299036 & 1.32 & 1.24 & 0.11 \\ 
00013299037 & 0.98 & 1.04 & 0.08 \\ 
00013299038 & 0.54 & 1.95 & 0.05 \\ 
00013299039 & 1.23 & 1.89 & 0.30 \\ 
00013299040 & 0.53 & 1.05 & 0.04 \\ 
00013299041 & 0.78 & 0.86 & 0.02 \\ 
00013299042 & 0.84 & 0.36 & 0.02 \\ 
00013299043 & 2.45 & 0.21 & 0.01 \\ 
00013299044 & 1.00 & 0.14 & 0.01 \\ 
00013299045 & 0.85 & 0.87 & 0.01 \\ 
00013299046 & 0.94 & 1.13 & 0.02 \\ 
00013299047 & 2.78 & 2.08 & 0.02 \\ 
00013299048 & 3.08 & 1.85 & 0.02 \\ 
00013299049 & 0.57 & 2.24 & 0.02 \\ 
00013299050 & 0.83 & 0.69 & 0.02 \\ 
00013299051 & 1.04 & 0.16 & 0.01 \\ 
00013299052 & 0.87 & 0.07 & 0.01 \\ 
00013299053 & 0.99 & 1.29 & 0.02 \\ 
00013299054 & 2.99 & 1.32 & 0.02 \\ 
00013299055 & 0.83 & 1.32 & 0.03 \\ 
00013299056 & 0.48 & 1.92 & 0.02 \\ 
00013299057 & 0.93 & 1.8 & 0.02 \\
00013299058 & 1.86 & 2.02 & 0.02 \\
\multicolumn{4}{|c|}{\textit{NuSTAR}/FPMA} \\
90701342002 & 55.78 & 3.94 & 0.07 \\
\multicolumn{4}{|c|}{\textit{NuSTAR}/FPMB} \\
90701342002 & 55.37 & 3.91 & 0.08 \\
\multicolumn{4}{|c|}{\textit{SRG}/ART-XC} \\
 & 0.55 & 0.43 & 0.02 \\
\hline 
\end{tabular} \\
\footnotesize{\blue{$^{\rm a}$} $10^{-1}$ cnt~s$^{-1}$ (total source region count rate)}\\
\footnotesize{\blue{$^{\rm b}$} $10^{-1}$ cnt~s$^{-1}$ (background count rate renormalised to the source region)}\\
\end{table}

\section{Results}

Using the available data described above we were able to study temporal and spectral properties of \eras\ at different time scales. Fig.~\ref{fig:lightcurve} shows a light curve of the source based on observations by the \textit{Swift}/XRT, \textit{NuSTAR}, and \textit{SRG}/ART-XC instruments between 59562 and 59722 MJD, clearly demonstrating an enhanced (about one order of magnitude) flux compared to the discovery value. The source is demonstrating a sinusoidal variability of the flux, which may point to the orbital periodicity in the system (for details see below).
About one year before the considered outburst, in the end of 2019--beginning of 2020 \eras\ was observed with \textit{SRG}/eROSITA and \textit{Swift}/XRT  \citep{Haberl2020}. The measured luminosities at that time fall into a narrow range of values around $0.5-1.0\times10^{36}$~\ergs\ (see grey shaded area at Fig.~\ref{fig:lightcurve}), which can be associated with the low state luminosity of the source. During the entire observation campaign, the source flux does not fall below this area.

\subsection{Timing analysis}

\begin{figure}
\centering
\includegraphics[width=0.99\columnwidth]{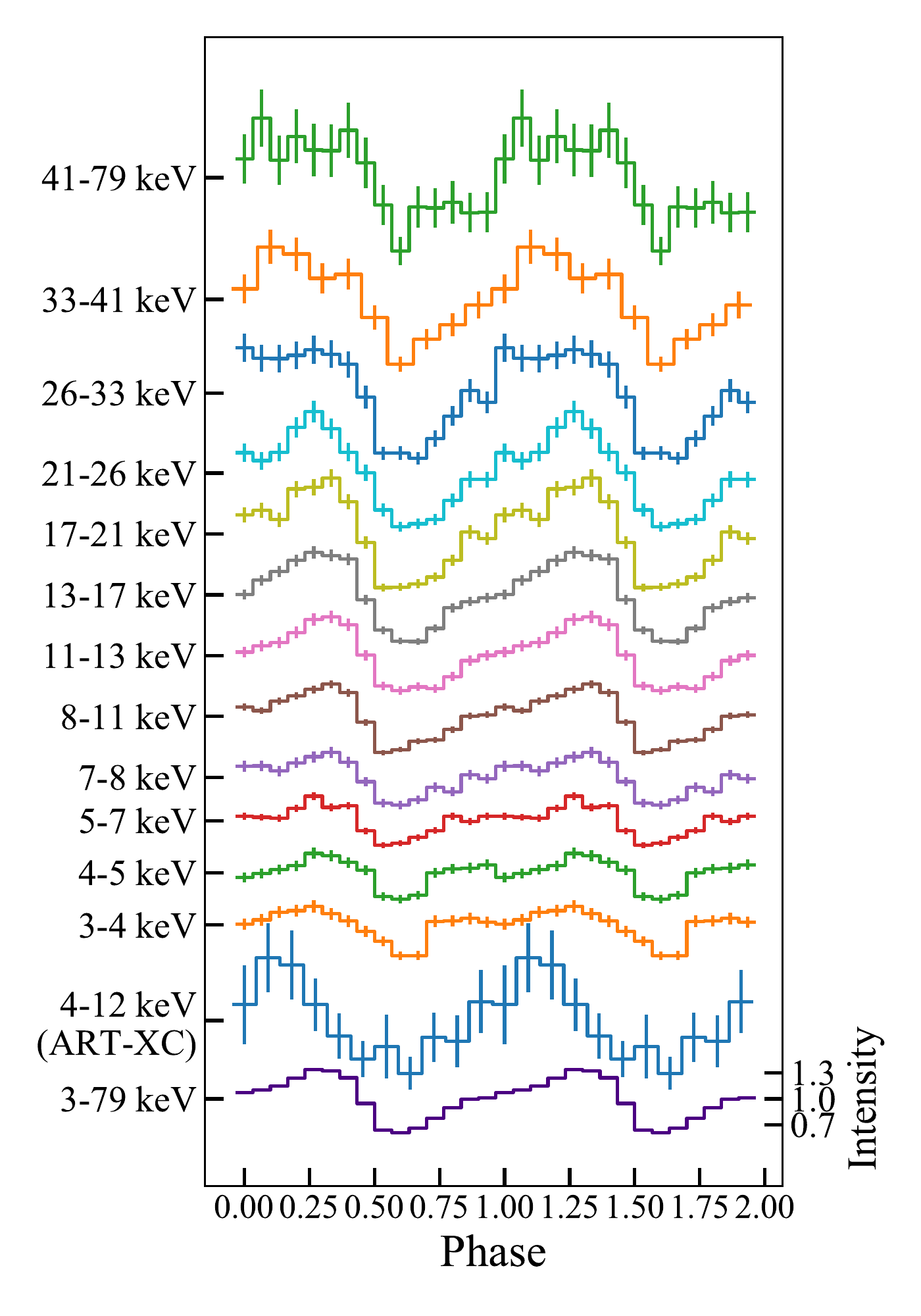}
\caption{\eras\ pulse profile as a function of the energy based on the \textit{NuSTAR} and \textit{SRG}/ART-XC data. Pulse profiles are repeated at phases 1.0-2.0 and spaced along the y-axis for the better visualization convenience. Intensities for each profile were normalized by the profile's average intensity. Zero phase values are different for \textit{NuSTAR} and \textit{SRG}/ART-XC.}
\label{fig:profiles}
\end{figure}

Preliminary results of the timing analysis performed on a part of the \textit{NuSTAR} data were published by  \cite{Haberl2021}. To refine the period of pulsations we used the full set of the \textit{NuSTAR} data in the energy range 3-79~keV.
%and applied the {\sc efsearch} utility. As the result we determined the period with the better accuracy than before $P_{\rm spin} = 40.5773 \pm 0.0001$~s. The pulse period uncertainty was determined using the epoch folding from $10^3$ simulated light curves  \citep[see][for details]{Boldin2013}. 
For that we used method based on a linear approximation of pulse arrival times \citep[see e.g.][]{Deeter1981}, resulted in the value $P_{\rm spin} = 40.5781 \pm 0.0004$~s. The spin period value was not corrected for binary motion, since the orbital parameters of the system are unknown except for the orbital period estimated in this work (see below). The {\it NuSTAR} wide energy range 3-79~keV, combined with high count statistics  allowed us to study the dependence of the \eras\ pulse profile on the energy (Fig.~\ref{fig:profiles}). Pulse profiles in different energy ranges were obtained by folding energy-resolved light curves with the measured period $P_{\rm spin}$ and normalized by the average intensity in a given energy band. Formally, we do not register pulsations from \eras\ in the ART-XC data due to the short exposure and insufficient statistics. Nevertheless, if we fold the light curve obtained by ART-XC in the 4-12 keV energy range with the period measured above, then the corresponding pulse profile will be generally similar to that obtained from the \textit{NuSTAR} data (see Fig.~\ref{fig:profiles}). 
Pulse profiles in all bands have a broad single-peaked profile with asymmetric wings. The intensity of the right wing relative to the left one increases with the energy. 

%======================================================
%======================================================

\begin{figure}
\centering
\includegraphics[width=0.99\columnwidth]{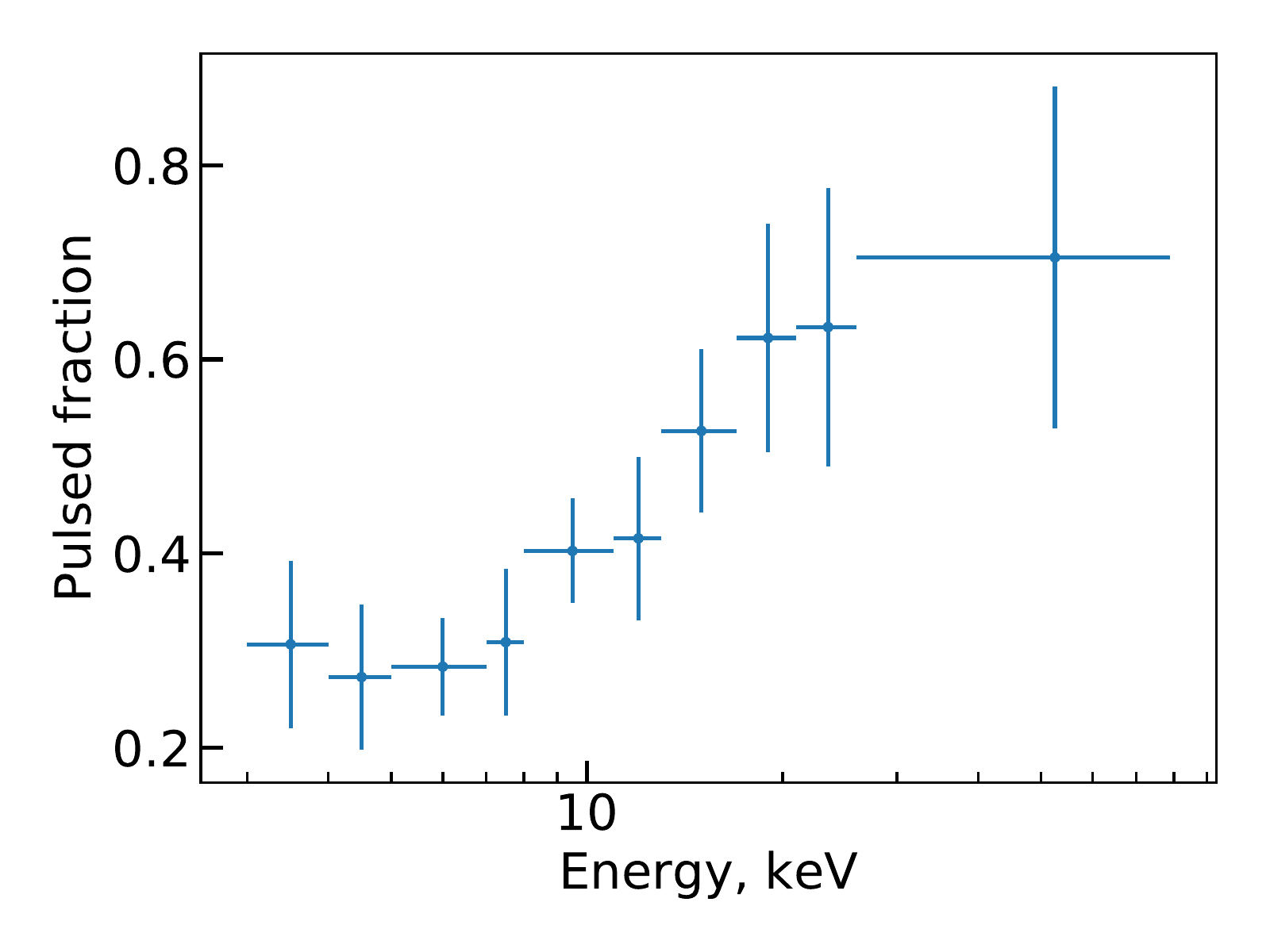}
\caption{Pulsed fraction of \eras\ as a function of the energy based on the \textit{NuSTAR} data.}
\label{fig:pulsed}
\end{figure}

We also studied a dependence of the pulsed fraction (PF)\footnote{PF is defined as [max (rate) - min (rate)] / [max (rate) + min (rate)], where max (rate) and min (rate) are maximum and minimum count rates in the profile, respectively.} on the energy (Fig.~\ref{fig:pulsed}).
The PF was calculated from the energy-resolved pulse profile with 15 phase bins. It demonstrates a monotonic increase in the entire energy range with the lowest value $\sim$ 30 per cent, that is typical for most bright XRPs \citep[][]{LutovinovTsygankov2009}.

Based on the conducted long-term monitoring of \eras, we were able to estimate the orbital period of the system. From the approximation of the light curve by a sinusoid we obtained a value of a possible orbital period of $38.0\pm0.1$ days (see Fig.~\ref{fig:lightcurve}). At the moment, a more accurate and detailed determination of the orbital parameters is not possible and additional observations are required.

%======================================================
%======================================================

\subsection{Spectral analysis}
\label{sec:spectrum}
\begin{figure}
\centering
\includegraphics[width=0.99\columnwidth]{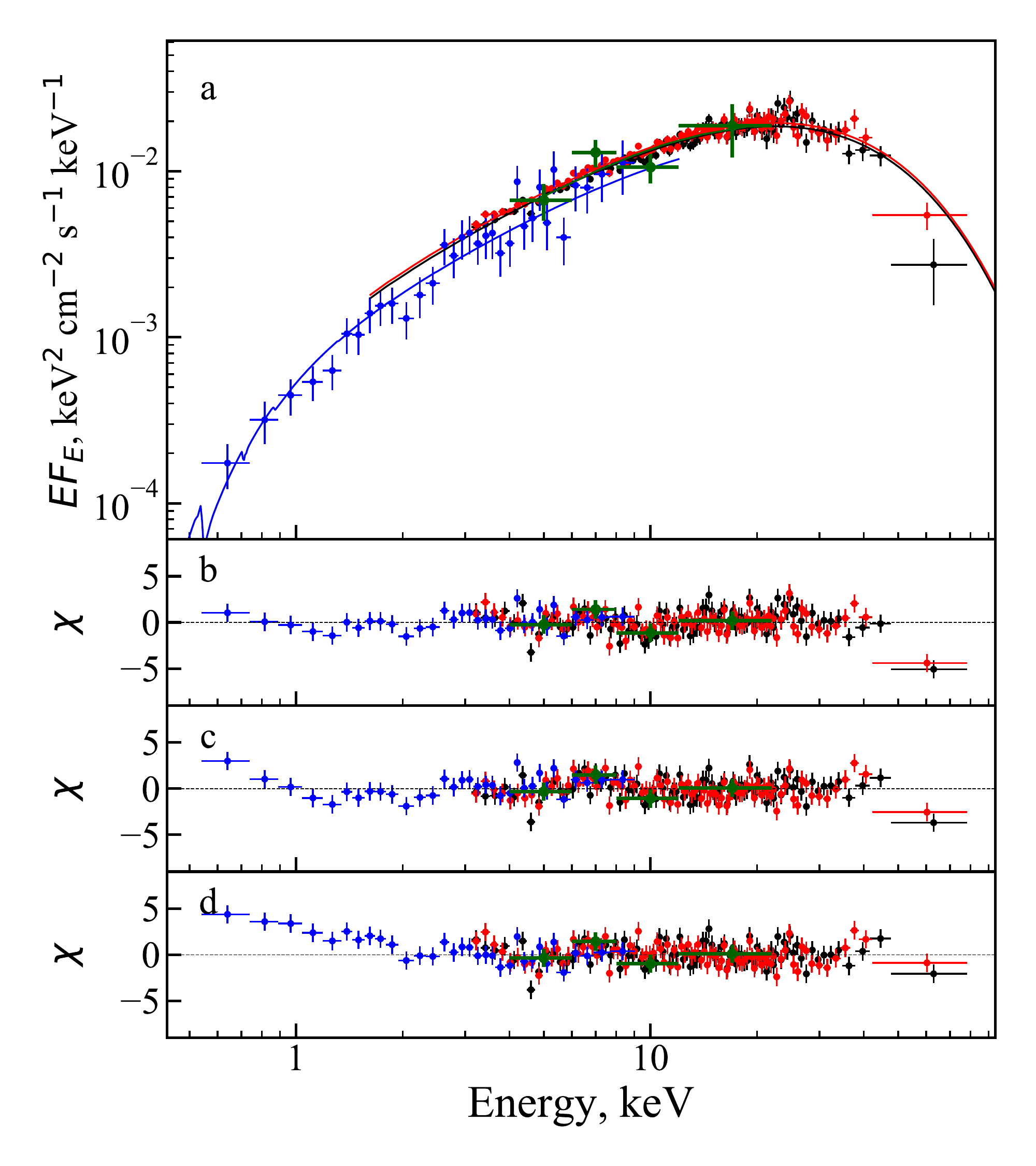}
\caption{Unfolded spectrum of \eras\ and its approximation with the model {\sc const $\times$ tbabs $\times$ cutoffpl} (solid lines in panel a). Red and black dots are for the FPMA and FPMB telescopes of the {\it NuSTAR} observatory, blue for the {\it Swift}/XRT telescope and dark green for the {\it SRG}/ART-XC telescope. 
The bottom three panels show the residuals for different continuum models: {\sc cutoffpl} (panel b), {\sc po $\times$ highecut} (panel c), {\sc compTT} (panel d).} 
\label{pic:avg_spectrum}
\end{figure}

To study the spectral properties of \eras\ in a wide energy range we jointly approximated data from the FPMA, FPMB, XRT and ART-XC telescopes. To get better statistics at a low-energy part of the source spectrum, we decided to use the averaged spectrum of several \textit{Swift}/XRT observations, which did not differ significantly in the flux (ObsID 00013299003, 00013299004, 00013299005, 00013299006, 00013299007), covering MJD 59568-59575  
The resulting phase averaged energy spectrum of \eras\ is shown in Fig.~\ref{pic:avg_spectrum}.
In order to take into account the non-simultaneity of observations as well as possible inaccuracies in calibrations between FPMA, FPMB, XRT and ART-XC, the cross-calibration multiplicative factors were used (the {\sc const} model in {\sc xspec}). All models were also modified by the photoabsorption model using the {\sc tbabs} component with the abundances from \citet{Wilms2000}.

%======================================================
%======================================================

%======================================================
\begin{table*}
\caption{Results of approximation of the \eras\ spectrum  by various continuum models. 
}
\begin{center}
\begin{tabular}{lccc} 
 \hline
 Parameter &{\sc cutoffpl} &  {\sc po $\times$ highecut $\times$ gabs}
 & {\sc compTT} \\
  \hline
 ${\rm const}_{\rm FPMA}$ & 1.00 (frozen)  &  1.00 (frozen) &  1.00 (frozen)  \\ 
 ${\rm const}_{\rm FPMB}$ & $1.05 \pm 0.01$ &  $1.05 \pm 0.01$ &  $1.05 \pm 0.01$ \\ 
 ${\rm const}_{\rm XRT}$ &  $0.79 \pm 0.04$  &  $0.73 \pm 0.04$ &  $0.95 \pm 0.05$  \\
 ${\rm const}_{\rm ART-XC}$ &  $1.0 \pm 0.1$  &  $1.1 \pm 0.1$ &  $1.0 \pm 0.1$  \\
 $N_{\rm H}$, $10^{22}$~cm$^{-2}$ & $0.27\pm0.07$ & $0.6\pm0.1$ & $0.25\pm0.25$ \\
 $E_{\rm cut}$,~keV &  &$19.5\pm 0.8$ &   \\
 $E_{\rm fold}$,~keV & $17.1\pm0.6$ & $17\pm1$ & \\ 
 $\Gamma$ & $0.68\pm0.03$ & $1.16\pm0.02$ &  \\ 
 $T_0$,~keV & &  & $0.87\pm0.04$\\ 
 $T$,~keV & & &$7.0\pm0.1$   \\ 
 $\tau$ & & &$5.7\pm0.1$   \\
 ${\rm Norm}_{\rm continuum}$, ph~keV$^{-1}$~s$^{-1}$~cm$^{-2}$  & $(1.14\pm0.04)\times10^{-3}$ &  $(1.87\pm0.06)\times10^{-3}$ & $(0.66\pm0.02)\times10^{-3}$   \\
 
 $E_{\rm smoothgabs}$,~keV &  &$19.5$ (=$E_{\rm cut}$) & \\ 
 $\sigma_{\rm smoothgabs}$,~keV &  &$1.95$ (=$0.1 \times E_{\rm cut}$) & \\
  $\tau_{\rm smoothgabs}$ &  &$0.20\pm0.05$ & \\
 Flux$_{0.5-79 \rm keV}$, \flux & $(6.56\pm0.09)\times10^{-11}$ & $(6.52\pm0.09)\times10^{-11}$ & $(6.19\pm0.07)\times10^{-11}$ \\
 Luminosity$_{0.5-79 \rm keV}$, \lum & $(1.96\pm0.03)\times10^{37}$ & $(1.95\pm0.03)\times10^{37}$ & $(1.85\pm0.02)\times10^{37}$ \\
  W-statistic/d.o.f. & 684/680 & 681/678 & 689/679 \\
 \hline
\end{tabular}
\label{table:spec_params}
\end{center}
\end{table*}
%======================================================
\begin{figure}
    \centering
 	\includegraphics[width=0.8\columnwidth]{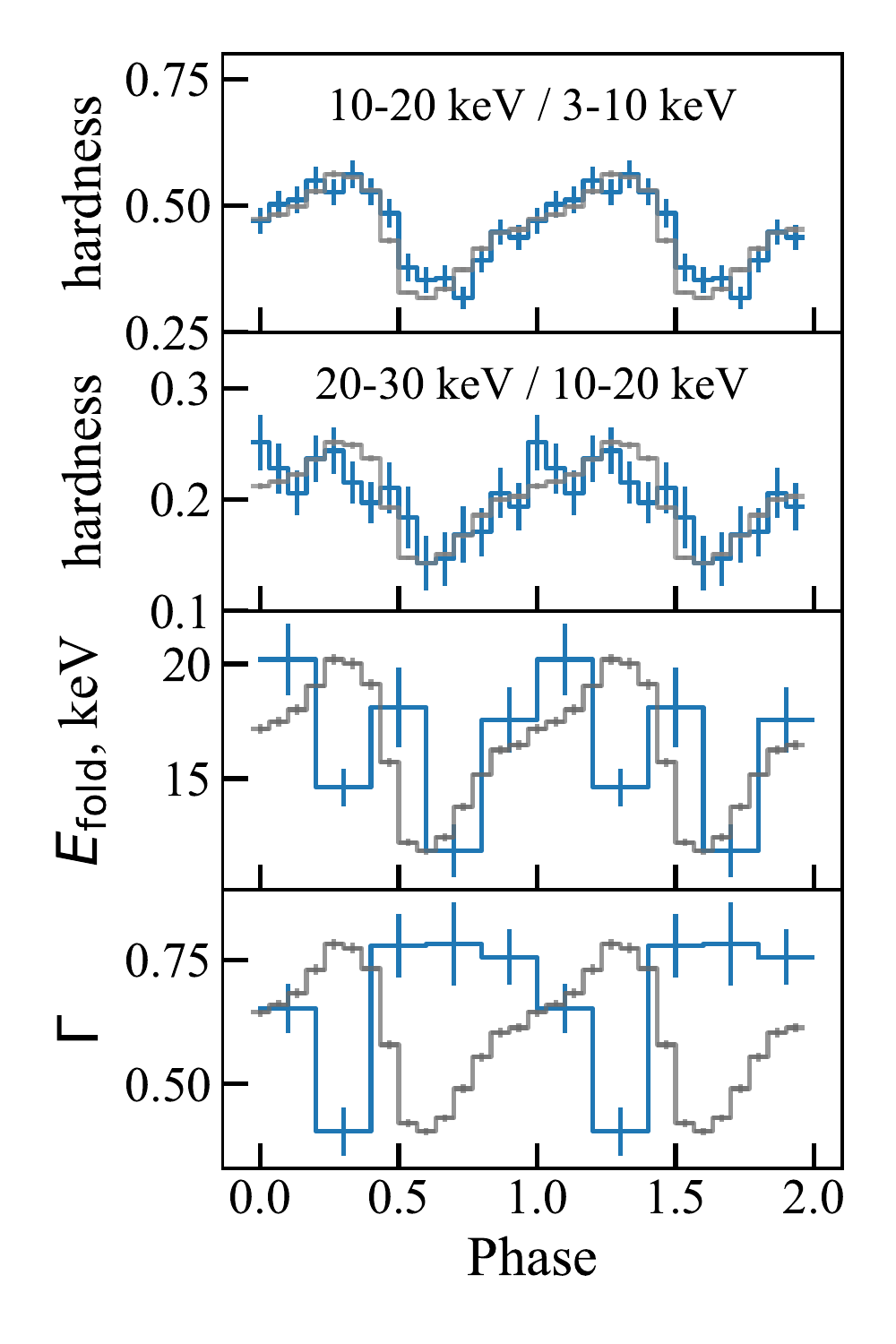}
 	\caption{The hardness ratio for \eras\ pulse profiles and  the spectral model parameters as a function of the spin phase. The averaged pulse profile in a wide energy range (3-79 keV) is superimposed in gray for visual comparison. Hardness is defined as the ratio of unnormalized pulse profiles in the corresponding energy bands (10-20~keV / 3-10~keV and 20-30~keV / 10-20~keV).}
	\label{fig:phres_spec}
\end{figure}
%======================================================

To approximate the spectrum, we used the cutoff power-law continuum models {\sc cutoffpl} and {\sc powerlaw $\times$ highecut} as well as the Comptonized radiation model {\sc compTT} \citep{Titarchuk1994} from the {\sc xspec} package. To smooth out the discontinuity in the {\sc powerlaw $\times$ highecut} continuum that forms artificial absorption-like residuals near the cutoff energy $E_{\rm cut}$, we have added a smoothing Gaussian absorption line {\sc gabs} at energy $E_{\rm smoothgabs}$ = $E_{\rm cut}$, with the width of $\sigma_{\rm smoothgabs}$ = 0.1 $E_{\rm cut}$ and the optical depth $\tau_{\rm smoothgabs}$ \citep[see][for details] {Coburn2002}.

The source spectrum has a typical shape for XRPs \citep{Coburn2002, Filippova2005}. All three above mentioned models describe it with similar quality, however cutoff power-law continuum models make it slightly better (see Table~\ref{table:spec_params}). 
Note that there is some spread in the value of a neutral hydrogen column density $N_{\rm H}$ depending on the selected continuum model and it is poorly defined in the model with the {\sc compTT} continuum. The value of $N_{\rm H}$ from the {\sc const $\times$ tbabs $\times$ cutoffpl} model is mostly close to the absorption value in the source direction $1.18\times10^{21}$ cm$^{-2}$ obtained by \citet{HI4PI2016}. Some excess of the value measured in X-rays is probably due to an additional absorption in the binary system. None of the continuum models revealed a significant presence of the Fe K$\alpha$ line. The $3\sigma$ upper limit flux for an narrow ($\sigma=0.1$~keV) iron line at 6.4 keV is $F_{\rm iron} = 1.2\times10^{-5} \text{ photons cm}^{-2}\text{ s}^{-1}$, which corresponds to an equivalent width of 0.05~keV. In the further analysis we will use the exponential cutoff power-law model {\sc const $\times$ tbabs $\times$ cutoffpl} as it has the smallest number of parameters. 

Thanks to the high counting statistics we were able to study an evolution of the source spectral parameters over the pulse phase using a pulse phase-resolved spectroscopy. The \textit{NuSTAR} spectral data were divided into five evenly distributed phases, each of which was described by the model used to describe the phase-averaged spectrum: {\sc const $\times$ tbabs $\times$ cutoffpl} with the $N_{\rm H}$ value fixed at 0.27.
Despite the fact that the model is phenomenological and makes it difficult to draw unambiguous conclusions about the ongoing physical processes, one can notice some characteristic features of the evolution of spectral parameters with a rotation phase.
A remarkable correlation (linear cross-correlation coefficient $\simeq 0.93$) between the energy flux and the hardness ratio over the pulse was revealed in our analysis (see Fig.~\ref{fig:phres_spec}). The spectrum hardness is defined as the ratio of non-normalized count rate pulse profiles in two adjacent energy bands. There is also some correlation between the folding energy $E_{\rm fold}$ and the pulse profile intensity, and an anti-correlation between the photon index $\Gamma$ and the pulse profile intensity. The latter one is generally agreed with the above mentioned correlation of the hardness with the pulse profile intensity.

\section{Discussion}

In this paper, we investigated properties of the recently discovered Be/X-ray binary \eras.  Below we discuss our observational results in framework of theoretical models of the accretion onto the magnetised NSs.

\subsection{\eras\ on the Corbet diagram}
Conventionally, it is believed that the main factor influencing the observed properties of pulsars is the angular momentum evolution of the NS as a result of accretion of matter \citep[see, e.g.,][]{2022arXiv220414185M}. This leads to an observational correlation between the pulsar spin period $P_{\rm spin}$ and the orbital period $P_{\rm orb}$ \citep[][]{Corbet1984, Corbet1985, Corbet1986}. This correlation clearly distinguishes between three main HMXB groups: Roche lobe-filling supergiants, wind accretion supergiants, and Be-HMXB.

\begin{figure}
\centering
\includegraphics[width=0.99\columnwidth]{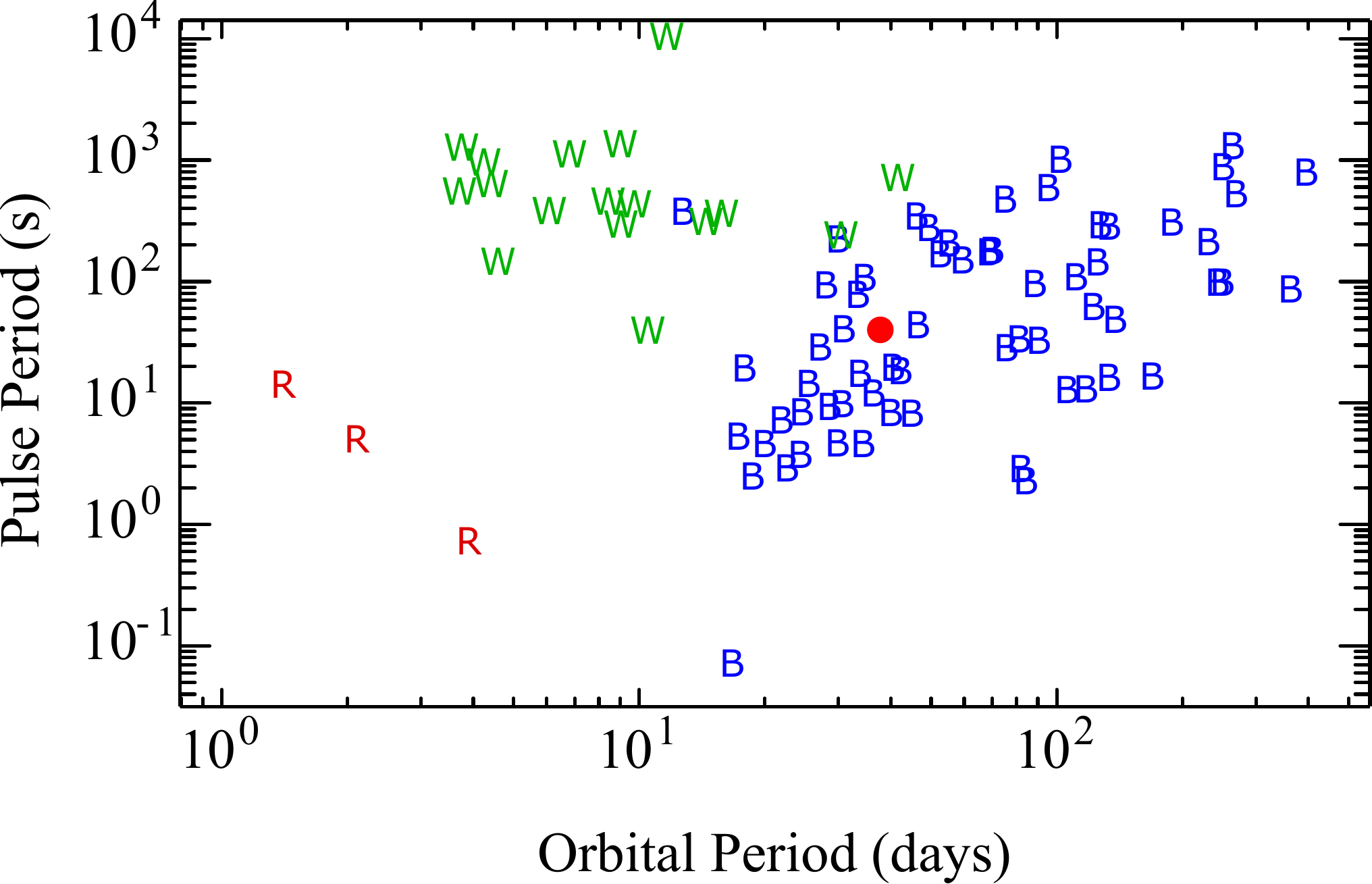}
\caption{Pulse period - orbital period diagram adopted from \citet{Corbet2017}. \eras\ is superimposed and marked as a red circle. "B” is for Be-star systems, “W” is for stellar wind acrretion, “R” is for Roche lobe-filling supergiants candidates.} 
\label{fig:diagram}
\end{figure}

The orbital 38-day and spin 40.6~s periods we measured allow us to study the position of \eras\ on the pulse period - orbital period Corbet diagram (see Fig~\ref{fig:diagram}). \eras\ is in the middle of Be-star systems on the Corbet diagram, which confirms its Be nature deduced from the SALT spectroscopy \citep{Haberl2020}.
%======================================================
%======================================================

\subsection{Magnetic field estimation}
\label{sec:magnetic}
In order to interpret the observed behaviour of a new XRP, some knowledge of the NS magnetic field is required.
The most accurate and only direct method for measuring XRP magnetic fields is to detect the cyclotron resonant scattering feature (CRSF) in its energy spectrum \citep[see][for a recent review]{Staubert2019,2022arXiv220414185M}. 
The analysis of the phase-averaged spectrum of \eras\ did not reveal the presence of CRSF in the energy range from 5 to 50 keV \citep[see ][for a brief description of the technique]{2005AstL...31...88T}.
It is important to note that in some cases, such spectral features appear only at some specific phases of the NS rotation  \citep{Molkov2019, Molkov2021}. However, in the case of \eras, the pulse phase-resolved spectroscopy did not reveal a phase-transient CRSF as well. An absence of the CRSF in the source energy spectrum allows us to constrain the XRP's magnetic field strength. For the energy range 5-50 keV and the canonical NS parameters ($M=1.4M_{\sun}$, $R=10^{6}$~cm), it points either to a low value $B<5\times10^{11}$~G or quite strong one $B>5\times10^{12}$~G. It is important to note that despite the absence of significant CRSF detection in the spectrum, an unusually shallow line can never be excluded.

If the direct measurement of the magnetic field strength using the CRSF energy is not possible, indirect methods can be applied.
When the accretion rate drops to very low values, the so-called propeller effect may take place \citep{Illarionov1975}, causing a sharp drop in the observed luminosity. As previously noted, \eras\ observations by \srg/eROSITA and {\it Swift}/XRT, performed in 2019-2020, when the source was not in the outburst, demonstrated relatively high luminosities, $1.0\times10^{36}$ and $4.8\times10^{35}$~\ergs, respectively \citep{Haberl2020}. Since such a luminosity is much higher than the typical quiescent luminosity of transient Be/XRPs \citep{Tsygankov2017}, it indicates the absence of a transition of the source to the propeller regime.  
The observed luminosity can  rather be associated with the accretion from the cold recombined accretion disc, observed in several XRPs with relatively long spin periods \citep{Tsygankov2017, Tsygankov2017_2, Tsygankov2019, Escorial2018, Reig2018}. 
In this case, the magnetic field can be estimated using formula (12) from \citet{Tsygankov2017_2} as $B\sim10^{13}$~G. This estimate should be treated with caution, since it was obtained under the assumption that accretion in the low state occurs from the “cold” recombined accretion disc. In the case of another accretion mechanism in the low state, this model is not applicable.

The NS magnetic field can also be indirectly estimated from an evolution of the pulsar's spin period. Measuring the spin-up and spin-down rates as a function of the mass accretion rate allows to study the interaction of the accreting plasma with the magnetosphere and estimate the magnetic field \citep[see, e.g.,][]{GhoshLamb1979b}. Unfortunately, the absence of a long observational history and, consequently, the absence of the orbital parameters of the system do not give us such a possibility yet.

At the same time, it is generally accepted that stably accreting XRPs, in the course of their evolution, come to an equilibrium of spin-up and spin-down torques, which is achieved at the so-called equilibrium period. As previously noted, \eras\ observations by \srg/eROSITA and {\it Swift}/XRT in 2019-2020 demonstrated quite high luminosities. These values differ only by a factor of 2 despite a significant gap of $\sim100$ days between observations, which allows us to assume the relative stability of the mass accretion rate in the low state
(see gray strip in Fig.~\ref{fig:lightcurve}). Although the source exhibits flux variability similar to the one discussed in this work, the dynamic range of the observed luminosity is still about one order of magnitude. 
This fact allows us to assume an equilibrium regime of accretion and roughly estimate the NS magnetic field considering the observed spin period as equilibrium one (see, e.g., Chapter 5 in \citealt{1992ans..book.....L}):
\begin{equation}
P_{\rm eq} \simeq 5.7\,\Lambda^{3/2}B_{12}^{6/7}L_{37}^{-3/7}M_{\rm s}^{-2/7}R_{6}^{15/7}~\text{s,}
\label{eq:period}
\end{equation}
where $\Lambda$ is the coefficient of proportionality between the Alfven radius and the radius of neutron star magnetosphere $R_{\rm m}$ (in the case of the accretion from gas pressure dominated disc, the coefficient is taken to be $\Lambda=0.5$), $B_{12}$ is the magnetic field strength $B$ at the NS surface in units of $10^{12}\,{\rm G}$, $L_{37}$ is the accretion luminosity $L$ in units of $10^{37}\,{\rm erg\,s^{-1}}$, $M_{\rm s}$ is the mass of a NS in units of solar masses $M_{\sun}$, and $R_{6}$ is the radius of a NS in units of $10^6\,{\rm cm}$.

The numerical factor in the right hand side of eq. (\ref{eq:period}) is determined by the critical fastness parameter $\omega_{\rm s}=({R_{\rm m}}/{R_{\rm cor}})^{3/2}$ (at which the total torque becomes zero), where $R_{\rm cor}$ is the corotation radius. 
The critical fastness parameter takes different values depending on the assumptions of the theoretical model. 
In particular, the model by \cite{GhoshLamb1979b} suggests $\omega_{\rm s}$=0.35, that corresponds to the factor value of 5.7 and magnetic field $B\sim1.5\times10^{13}$~G, whereas the model by \cite{Wang1995} suggests $\omega_{\rm s}$=0.95 (factor value is 2.1) resulting in $B\sim5\times10^{13}$~G. Both values are in agreement with estimates obtained from the accretion from the cold disc and point to the relatively strong magnetic field in \eras.
 However, such a magnetic field estimation should be treated with the caution since there is no sufficiently long and detailed history of observations of \eras\ in the quiescent state to claim the presence of the spin equilibrium confidently.

\subsection{Emission mechanism}

The pulse profile of \eras\ in all energy ranges demonstrates a single-peak structure, which indicates that the pulsar is probably turned towards the observer by one of its poles, while another one is practically invisible. As previously noted, there is a strong correlation between the energy flux and the hardness ratio over the pulse (see Sec.~\ref{sec:spectrum}). Such an unusually strong correlation may indicate the dominant nature of the pencil beam pattern in the NS radiation. As shown below, it can be caused, depending on the NS's magnetic field, either by the features of the energy release in the atmosphere or by the details of photons propagation above its surface.

If the magnetic field strength at the NS surface is extremely high, the critical luminosity is well above the observed luminosity level of $\sim 2\times 10^{37}\,{\rm erg\,s^{-1}}$ 
\citep{1976MNRAS.175..395B,2015MNRAS.447.1847M}. Then the correlation between the flux and hardness ratio may be a result of temperature structure in the atmosphere and dependence of Compton scattering cross section on the photon momentum direction. The beam pattern produced by the hot spots at the stellar surface is close to the pencil beam. Most photons leaving the surface do not experience scattering by the accreting matter. Photons propagating closer to the normal to the NS surface are originated from deeper layers in the atmosphere due to the geometrical reason and strong dependence of the scattering cross section on the angle between the magnetic field direction and photon momentum \citep{1975A&A....42..311B}. The deeper atmospheric layers are expected to be hotter than the upper ones at the considered mass accretion rates. As a result, the photons propagating close to the perpendicular to the NS surface experience a thermal Comptonization by the electron gas of the higher temperature and, thus, have harder spectra. This scenario was utilised earlier by \citet{2008A&A...491..833K} to explain a similar correlation between the flux and hardness ratio in XRP EXO 2030+375.

In the case of the moderate magnetic field strength, the observed luminosity $\sim 2\times 10^{37}\,{\rm erg\,s^{-1}}$ is close to the critical value. In this case, a significant fraction of X-ray photons leaving the NS surface experience a Compton scattering in the accretion channel above the hot spots, and the bulk Comptonization becomes important \citep{2007ApJ...654..435B}. The Compton scattering by the accreting plasma is strongly anisotropic due to the high free-fall velocity of the accretion flow. Momenta of the majority of scattered photons are directed close to the direction of the accretion flow motion. The closer the direction of the scattered photon momenta is to the direction of the accretion flow, the energy of the scattered photons is higher. As a result, one would expect that the direction of predominant photon scattering is correlated with the direction of the hard emission. The photons experienced scattering in the accretion channel can undergo a following reflection from the NS surface, which makes the spectra even harder \citep{2015MNRAS.452.1601P} and conserves a correlation between the predominant direction of the photon motion and hardness ratio. Thus, one would expect a correlation between the flux and hardness ratio during the pulse period at accretion luminosities close to the critical value.

Based on the estimates of the magnetic field strength obtained above ({\it several} $10^{13}$~G), we consider the second scenario as the most probable one. We emphasize that even in this situation, the pulsar's beam function is expected to have a pencil configuration due to the reflection of scattered emission from the NS surface.

\section{Conclusion}

This paper presents results of the first spectral and temporal analysis in a wide energy range of recently discovered Be/X-ray binary \eras. The investigation is based on the data from the \textit{SRG}, \textit{NuSTAR}, and \textit{Swift} observatories obtained during the December 2021 -- May 2022 monitoring of the source. Timing analysis revealed pulsations with a period of  $40.5781$~s and a monotonic increase in PF over the entire energy range. Based on the long-term \textit{Swift}/XRT monitoring of the system, we obtained an estimate for the system's orbital period of 38 days. The \eras\ spectrum has the form of a power law with an exponential cutoff at high energies. We did not detect the presence of CRSF both in the phase-averaged and phase-resolved spectra. Based on indirect methods, estimates were given for the magnetic field about {\it several} $10^{13}$~G. The hardness ratio and pulse profile intensity demonstrate a strong correlation, which we interpret as a consequence of the bulk Comptonization of the emission by the accretion flow. 

The high sensitivity of the instruments of the \textit{SRG} observatory and the possibility of surveying large sky areas allow it to discover many faint transient or quasi-permanent objects, including X-ray pulsars, both in our Galaxy \citep[see, e.g.,][]{ 2021arXiv210614539D, 2021arXiv210705587L} and neighboring galaxies, primarily in the Large Magellanic Cloud \citep[see, e.g.,][]{2021MNRAS.504..326M, 2022arXiv220300625H}. The latter is most likely due to the fact that the LMC is too large to be completely covered by the relatively small field of view of other sensitive instruments, like \textit{Chandra} or \textit{XMM-Newton}. The \textit{SRG} observatory survey strategy, as well as deep observations during the CalPV phase, have significantly increased the coverage of this region of the sky with a high sensitivity. This has already led to the discovery of several new pulsars, including \eras, which is the subject of this paper. It is noteworthy that most of the new pulsars, discovered by \textit{SRG} both in our Galaxy and in the LMC, reside in systems with Be stars.

%%%%%%%%%%%%%%%%%%%%%%%%%%%%%%%%%%%%%%%%%%%%%%%%%%%%%%%%%%%%%%%%%%%%%%%%%%%%%%
%% Acknowledgments                                                         %%
%%%%%%%%%%%%%%%%%%%%%%%%%%%%%%%%%%%%%%%%%%%%%%%%%%%%%%%%%%%%%%%%%%%%%%%%%%%%%%
\section*{Acknowledgements}
This work is based on data from {\it Mikhail Pavlinsky} ART-XC X-ray telescope aboard the SRG observatory. The SRG observatory was built by Roskosmos in the interests of the Russian Academy of Sciences represented by its Space Research Institute (IKI) in the framework of the Russian Federal Space Program, with the participation of the Deutsches Zentrum für Luft- und Raumfahrt (DLR). The SRG spacecraft was designed, built, launched and is operated by the Lavochkin Association and its subcontractors. The science data are downlinked via the Deep Space Network Antennae in Bear Lakes, Ussurijsk, and Baykonur, funded by Roskosmos. The ART-XC team thank the Russian Space Agency, Russian Academy of Sciences and State Corporation Rosatom for the support of the SRG project and ART-XC telescope and the Lavochkin Association (NPOL) with partners for the creation and operation of the SRG spacecraft (Navigator). We are grateful to the \textit{NuSTAR} team for approving and rapid scheduling of the follow-up observation. This work is based on data from {\it Mikhail Pavlinsky} ART-XC X-ray telescope aboard the SRG observatory. We are grateful to the \textit{Swift} team for approving and rapid scheduling of the monitoring campaign. This work made use of data supplied by the UK \textit{Swift} Science Data Centre at the University of Leicester and data obtained with \textit{NuSTAR} mission, a project led by Caltech, funded by NASA and managed by JPL. This research also has made use of the NuSTAR Data Analysis Software ({\sc NUSTARDAS}) jointly developed by the ASI Science Data Center (ASDC, Italy) and Caltech. This research has made use of data and software provided by the High Energy Astrophysics Science Archive Research Center (HEASARC), which is a service of the Astrophysics Science Division at NASA/GSFC and the High Energy Astrophysics Division of the Smithsonian Astrophysical Observatory. This work was supported by the grant 19-12-00423 of the Russian Science Foundation. SST also acknowledges the support from the Academy of Finland travel grant 349373.

\section{Data availability}

\textit{NuSTAR} and \textit{Swift} data can be accessed from corresponding online archives. At this moment \textit{SRG}/ART-XC data have a private status. They will be open for the scientific community after a special decision of the Roscosmos. 
%%%%%%%%%%%%%%%%%%%%%%%%%%%%%%%%%%%%%%%%%%%%%%%%%%%%%%%%%%%%%%%%%%%%%%%%%%%%%%
%% Bibliography                                                             %%
%%%%%%%%%%%%%%%%%%%%%%%%%%%%%%%%%%%%%%%%%%%%%%%%%%%%%%%%%%%%%%%%%%%%%%%%%%%%%%
 
\bibliographystyle{mnras}
\bibliography{allbib}

 % Don't change these lines
\bsp    % typesetting comment
\label{lastpage}
\end{document}